\documentclass[prl,showpacs,showkeys,twocolumn]{revtex4}
\usepackage{bm}
\usepackage{graphicx}
\begin{document}
\def\Barcelo{Barcel\'o}
\title{Towards the observation of Hawking radiation in Bose--Einstein condensates.}
\author{Carlos \Barcelo}
\email{carlos@hbar.wustl.edu}
\affiliation{Physics Department, Washington University, Saint
Louis, MO 63130--4899, USA}
\author{Stefano Liberati}
\email{liberati@physics.umd.edu}
\homepage{http://www2.physics.umd.edu/~liberati}
\thanks{Supported by the US NSF}
\affiliation{Physics Department, University of Maryland, 
College Park, MD 20742--4111, USA}
\author{Matt Visser}
\email{visser@kiwi.wustl.edu}
\homepage{http://www.physics.wustl.edu/~visser}
\thanks{Supported by the US DOE}
\affiliation{\centerline{Physics Department, Washington University, Saint
Louis MO 63130--4899, USA}\\}
\date{6 October 2001; \LaTeX-ed \today}
\bigskip
\begin{abstract}
Acoustic analogues of black holes (dumb holes) are generated when a
supersonic fluid flow entrains sound waves and forms a trapped region
from which sound cannot escape.  The surface of no return, the
acoustic horizon, is qualitatively very similar to the event horizon
of a general relativity black hole. In particular Hawking radiation (a
thermal bath of {\emph{phonons}} with temperature proportional to the
``surface gravity'') is expected to occur. In this note we consider
quasi-one-dimensional supersonic flow of a Bose--Einstein condensate
(BEC) in a Laval nozzle (converging-diverging nozzle), with a view to
finding which experimental settings could magnify this effect and
provide an observable signal. We identify an experimentally plausible
configuration with a Hawking temperature of order $70$ n K; to be
contrasted with a condensation temperature of the order of $90$ n K.
\end{abstract}
\pacs{04.40.-b; 04.60.-m; 11.10.-z; 45.20.-d; gr-qc/0110036}
\keywords{Laval nozzle, converging-diverging nozzle, supersonic flow, surface gravity}
\maketitle
\def\half{{1\over2}}
\def\L{{\mathcal L}}
\def\S{{\mathcal S}}
\def\d{{\mathrm{d}}}
\def\etal{{\emph{et al.}}}
\def\det{{\mathrm{det}}}
\def\tr{{\mathrm{tr}}}
\def\ie{{\emph{i.e.}}}
\def\eg{{\emph{e.g.}}}
\def\im{{\rm i}}
\def\bnabla{\mbox{\boldmath$\nabla$}}
\def\x{{\mathbf x}}
\def\aka{{\emph{aka}}}
\def\Choose#1#2{{#1 \choose #2}}
\def\etc{{\emph{etc.}}}
\def\Hospital{H\^opital}
\def\HRULE{{\bigskip\hrule\bigskip}}
\def\be{\begin{equation}}
\def\ee{\end{equation}}
\def\bea{\begin{eqnarray}}
\def\eea{\end{eqnarray}}

Acoustic analogues of black holes are formed by supersonic fluid
flow~\cite{unruh,visser}. The flow entrains sound waves and forms a
trapped region from which sound cannot escape.  The surface of no
return, the acoustic horizon, is qualitatively very similar to the
event horizon of a general relativity black hole; in particular
Hawking radiation (in this case a thermal bath of {\emph{phonons}}
with temperature proportional to the ``surface gravity'') is expected
to occur~\cite{unruh,visser}.  There are at least three physical
situations in which acoustic horizons are known to occur: Bondi--Hoyle
accretion~\cite{bondi-hoyle}, the Parker wind~\cite{parker} (coronal
outflow from a star), and supersonic wind tunnels. Recent improvements
in the creation and control of Bose--Einstein condensates
(see \eg,~\cite{Dalf99}) have lead to a growing interest in these
systems as experimental realizations of acoustic analogs of event
horizons.  In this note we consider supersonic flow of a BEC through a
Laval nozzle (converging-diverging nozzle) in a quasi-one-dimensional
approximation. We show that this geometry allows the existence of a
fluid flow with acoustic horizons without requiring any special
external potential, and we study this flow with a view to finding
situations in which the Hawking effect is large.  We present simple
physical estimates for the ``surface gravity'' and Hawking
temperature~\cite{footnote}. 

While writing up this paper, we have found that fluid flow in a Laval
nozzle geometry has also been considered in reference~\cite{sakagami}.
The occurrence of the same Laval nozzle geometry is the only
significant point of overlap of our work with~\cite{sakagami}, as that
paper only deals with a classical effect, related to the Hawking
effect, but does not consider the quantum physics of Hawking radiation
itself.  (See~\cite{padmanabhan} for a previous discussion along these
lines.)

\emph{Laval nozzle:} 
A general problem with the realization of acoustic horizons is that
most of the background fluid flows so far studied seem to require very
special fine-tuned forms for the external potential in order to be
realized (see \eg, the Schwarzschild-like geometry in
reference~\cite{unexpected}). In this respect a possible improvement
toward the realizability of acoustic horizons is the construction of a
trap which ``geometrically constrains'' the flow in such a way as to
replace the need for a special external potential. An example of such
a geometry is the so called Laval nozzle. In particular we shall
consider a pair of Laval nozzles; this provides a system which
includes a region of supersonic flow bounded between two subsonic
regions.

\begin{figure}[htbp]
\vbox{
\hfil
\scalebox{1.00}{\includegraphics{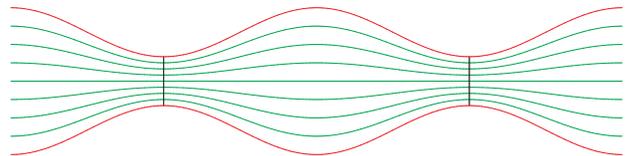}}
\hfil
}
\bigskip
\caption[Pair of Laval nozzles]{%
{\sl A pair of Laval nozzles: The second constriction is used to bring
the fluid flow back to subsonic velocities. 
\smallskip}}
\end{figure}
Consider such a nozzle pointing along the $z$ axis.  Let the cross
sectional area be denoted $A(z)$.  We apply, with appropriate
modifications and simplifications, the calculation
of~\cite{unexpected}. The crucial approximation is that transverse
velocities (in the $x$ and $y$ directions) are small with respect to
velocity along the $z$ axis. Then, assuming steady flow, we can write
the continuity equation in the form
\be
\rho(z) \; A(z) \; v(z) = J; \qquad J=\hbox{constant.}
\label{eq:cont}
\ee
The Euler equation (including for the moment possible external body
forces $\d \Phi/\d z$, and internal viscous friction $f_v$) reduces to
\be
\rho \; v \; {\d v\over \d z} = 
- {\d p\over\d z} - \rho \; {\d\Phi\over\d z} + f_v.
\label{eq:eul}
\ee
Finally, we assume a barotropic equation of state $\rho = \rho(p)$,
and define $X' = \d X/\d z$.  Then continuity implies
\be
\rho' = - \rho \; {(A v)'\over(A v)} = 
-\rho \left[ {A'\over A} + {v'\over v}\right] = 
-\rho  \left[ {A'\over A} + {a\over v^2}\right],
\ee
while Euler implies
\be
\rho \;a = - {\d p\over\d\rho} \; \rho'  - \rho \; \Phi' + f_v.
\ee
Defining the speed of sound by $c^2 = \d p/\d\rho$, and eliminating
$\rho'$ between these two equations yields a form of the well-known
``nozzle equation''
\be
a =  - {v^2\over c^2 - v^2} 
\left( c^2 \left[ {A'\over A} \right]   - \Phi' + {f_v\over\rho} \right)
\label{E:nozzle}
\ee
The presence of the factor $ c^2 - v^2$ in the denominator is crucial
and leads to several interesting physical effects. For instance, if
the physical acceleration is to be finite at the acoustic horizon, we
need
\be
c^2 \left[ {A'\over A} \right]   - \Phi' + {f_v\over\rho} \;\to\; 0.
\label{E:fine-tune}
\ee
This is a fine-tuning condition relating the shape of the nozzle (the
amount of convergence/divergence at the horizon) to the external body
force and specific friction. Experience with wind tunnels has shown
that the flow will attempt to self-adjust (in particular, the location
of the acoustic horizon will self-adjust) so as to satisfy this fine
tuning. We could now calculate the acceleration of the fluid at the
acoustic horizon by adopting L'\Hospital's rule. However, it is more
useful to consider the ``surface gravity'' defined by the limit of the
quantity~\cite{visser}
\be
g = {1\over2}{d(c^2-v^2)\over \d z}
\ee
It is this combination $g$, rather than the physical acceleration of
the fluid $a$, that more closely tracks the notion of ``surface
gravity'', and it is the limit of this quantity as one approaches the
acoustic horizon that enters into the Hawking radiation
calculation~\cite{essential}.  To calculate the limit $g_H$ we need to
use
\bea
(c^2)' &\equiv& {\d^2p\over\d\rho^2} \; \rho' =   
-\rho\; {\d^2p\over\d\rho^2} \; {(A\;v)'\over A\;v} 
\nonumber
\\
&\to& 
- \rho_H  \; {\d^2p\over\d\rho^2} \bigg|_{H}\; 
\left[{A'_H\over A_H}+{a_H\over c_H^2} \right].
\label{E:c2'}
\eea
This implies that the fine-tuning (\ref{E:fine-tune})
used to keep $a$ finite at the acoustic horizon will also keep $g$
finite there. In particular, we can use L'\Hospital's rule at the
acoustic horizon to evaluate
\be
g_H = -{\rho\over2} {\d^2 p\over \d\rho^2}{A'\over A}\bigg|_H
+
{\left[ c^2 + {\rho\over2} {\d^2 p\over \d\rho^2} \right] \left.
\left[c^2 {A'\over A}   -\Phi' + {f_v\over\rho} \right]'\right|_{H}
\over 2 g_H}
\ee 
This result, which we quote for its generality, is actually more
complicated than we really need.

\emph{Free flow:}
The Laval nozzle corresponds to the special case in which there is no
external body force $\Phi'=0$. Additionally, when considering a
superfluid flow, there is no viscous friction $f_v=0$.  Then the
nozzle equation (\ref{E:nozzle}) reduces to
\be
a = -{c^2 \; v^2\over c^2 - v^2}\left[ {A'\over A} \right].
\ee
Regularity now requires the much simpler fine-tuning condition that
$A'=0$ at the horizon. That is, the acoustic horizon occurs at a point
of minimum area; exactly the behaviour which is physically seen in a
Laval nozzle. (That a horizon cannot form at a maximum of the cross
sectional area is established below.)  Now apply the {L'\Hospital}
rule at the horizon (using the fact that at this point $A'=0$)
\be
a_H = {- c^4 A''/A \over (c^2)' - 2 a_H }\bigg|_{H}.
\ee
But from the specialization of (\ref{E:c2'}) we now see
\be
a_H^2 = {c^4 \; A''/A \over 2 +  \rho   (\d^2p/\d\rho^2)/c^2}\bigg|_{H}.
\ee
That is
\be
a_H = \pm {c^2\over\sqrt{2A}}\; \left.
\sqrt{A''\over1 + (1/2)  \rho [\d^2p/\d\rho^2]/c^2}\right|_{H}.
\ee
Very similar formulae hold for $g_H$: 
\be g_H^2 = + {1\over2} \left[ c^2 + {\rho\over2} {\d^2 p\over
\d\rho^2} \right]_H c^2 \left[ {A''_H\over A_H} \right], \ee
and so
\be
g_H = 
\pm
{c^2_H \over \sqrt{2A_H}} \; 
\left.\sqrt{ 1 + {\rho\over2c^2} {\d^2 p\over \d\rho^2}}\right|_H \; 
\sqrt{A''_H}.
\label{E:g_H}
\ee 
The first factor is of order $c^2_H/R$, with $R$ the minimum radius of
the nozzle, while the second and third factors are square roots of
dimensionless numbers. This is in accord with our intuition based on
dimensional analysis~\cite{visser,unexpected}. If $A''<0$,
corresponding to a maximum of the cross section, then $g_H$ is
imaginary which means no event horizon can form there. The two signs
$\pm$ correspond to either speeding up and slowing down as you cross
the horizon, both of these must occur at a minimum of the cross
sectional area $A''>0$. (If the flow accelerates at the horizon this
is a black hole horizon [future horizon]; if the flow decelerates
there it is a white hole horizon [past horizon]. See Figure 1.) If the
nozzle has a circular cross section, then the quantity $A''_H$ is
related to the longitudinal radius of curvature $R_{c}$ at the throat
of the nozzle, in fact $A_H'' = \pi {R / R_c}$.

\emph{Bose--Einstein condensate:}
The use of BECs as a working fluid for acoustic black holes has been
advocated by Garay \etal~\cite{garay} (see also~\cite{barcelo} for a
discussion of plausible extensions to that model). The present note
can be interpreted as a somewhat different approach to the same
physical problem, side-stepping the technical complications of the
Bogoliubov equations in favour of a more fluid dynamical point of
view. For a standard BEC
\be
c^2 = {\lambda \rho\over m}.
\ee
Then
\be
\rho \left[{\d^2p\over\d\rho^2}\right] = \rho {\d(c^2)\over\d\rho} = c^2,
\ee
while
\be
1 + {1\over2}  {\rho\over c^2} \left[{\d^2p\over\d\rho^2}\right] = {3\over2}.
\ee
So we have, rather simply
\be
a_H = \pm{c^2_H\over\sqrt{A_H}}\; \sqrt{A''_H/3}.
\ee
Similarly
\be
g_H = \pm{c^2_H \over \sqrt{A_H}} \; \sqrt{3A''_H/4}.
\ee 
This implies, at a black hole horizon [future horizon], a Hawking
temperature~\cite{unruh,visser,essential}
\be
k_{\rm B} T_H = {\hbar g_H\over2\pi c_H} = 
\hbar {c_H \over 2\pi \sqrt{A_H}} \; \sqrt{3A''_H\over4}.
\ee
Ignoring for now the issue of gray-body factors (they are a refinement
on the Hawking effect, not really an essential part of the physics),
the phonon spectrum peaks at
\be
\omega_{\mathrm{peak}} =  {c_H \over 2\pi \sqrt{A_H}} \; \sqrt{3A''_H\over4},
\ee
that is
\be
\lambda_{\mathrm{peak}} =4\pi^2 \sqrt{A_H} \; \sqrt{4\over3A''_H}.
\ee
This extremely simple result relates the Hawking emission to the
physical size of the constriction and a factor depending on the
flare-out at the narrowest point. Note that you cannot permit $A''_H$
to become large, since then you would violate the
quasi-one-dimensional approximation for the fluid flow that we have
been using in this note. (There is of course nothing physically wrong
with violating the quasi-one-dimensional approximation, it just means
the analysis becomes more complicated. In particular, if there is no
external body force and the viscous forces are zero then by slightly
adapting the analysis of~\cite{unexpected} the acoustic horizon [more
precisely the ergo-surface] is a minimal surface of zero extrinsic
curvature.)  The preceding argument suggests strongly that the best we
can realistically hope for is that the spectrum peaks at wavelength
\be
\lambda_{\mathrm{peak}} \approx \sqrt{A_H}.
\ee
You can (in principle) try to adjust the equation of state to make the
second factor in (\ref{E:g_H}) larger, but this is unlikely to be
technologically feasible.

(Note that this is the analog, in the context of acoustic black holes,
of the fact that the Hawking flux from general relativity black holes
is expected to peak at wavelengths near the physical diameter of the
black hole, its Schwarzschild radius --- up to numerical factors
depending on charge and angular momentum.)

\emph{Discussion:}
It is the fact that the peak wavelength of the Hawking radiation is of
order the physical dimensions of the system under consideration that
makes the effect so difficult to detect. In particular, in BECs it is
common to have a sound speed of order $6\;\hbox{mm/s}$.  If one then
chooses a nozzle of diameter about 1 micron, and a flare-out of
$A''_H\approx 1$, then $T_H \approx 7 \;\hbox{n K}$.  Compare this to
the condensation temperature required to form the BEC
\be
T_{\mathrm{condensate}} \approx 90 \;\hbox{n K}.
\ee
We see that in this situation the Hawking effect, although tiny, is at
least comparable in magnitude to other relevant temperature scales.
Moreover recent experiments indicate that it is likely that these
figures can be improved.  In particular, the scattering length for the
condensate can be tuned by making use of the so called Feshbach
resonance~\cite{feshbach}. This effect can be used to increment the
scattering length; factors of up to 100 have been experimentally
obtained~\cite{feshbach2}. Therefore the acoustic propagation speed,
which scales as the square root of the scattering length, could
thereby be enhanced by a factor up to 10. This suggests that it might
be experimentally possible to achieve $c_H\approx 6\;\hbox{cm/s}$, and
so 
\be
T_H \approx 70\;\hbox{n K}, 
\ee
which places us much closer to the condensation temperature.  The
speed of sound can also be enhanced by increasing the density of the
condensate (propagation speed scales as the square root of the
density). In all of these situations there is a trade-off: For fixed
nozzle geometry the Hawking temperature scales as the speed of sound,
so larger sound speed gives a bigger effect but conversely makes it
more difficult to set up the supersonic flow.

So if the Hawking effect can be experimentally realized in these
situations, it may be sufficiently large to disrupt the condensate
configuration, or even the condensate itself, providing in this way a
clear signal. Let us elaborate this point: The Garay {\etal}
analysis~\cite{garay} shows that it should be possible to create
classically stable BEC configurations with the presence of acoustic
horizons. In their particular analysis they found that these
classically stable configurations are surrounded by unstable regions;
(at this stage, we don't know how general that result is). Once the
system is engineered to be in a classically stable configuration one
can look for the purely quantum effect of Hawking emission.  The power
loss due to Hawking radiation would be:
\be
P = \sigma \;T_H^4 \;A_H =  {3 \; \hbar \;c_H^4\over 5120 \;\pi^2 \;A_H}  \; (A''_H)
^2.
\ee
Numerically (including the effect of the Feshbach resonance), the
emitted power is extremely small $P \approx 10^{-48} \hbox{ W}$, but
it should be noted that some sizable fraction of the Hawking phonons
will be sufficiently energetic to knock atoms out of the condensate
phase.

Additionally, we want to stress that the present analysis of the BEC
is purely ``hydrodynamic'' (superfluid approximation of the BEC), and
does not seek to deal with the ``quantum potential''(see
\eg,~\cite{Dalf99,barcelo}). The latter is responsible for the fact
that the dispersion relation for perturbations in the BEC is modified
at high momenta in such a way as to recover ``infinite'' propagation
speed (this is the so-called Bogoliubov dispersion
relation~\cite{barcelo,cpt01}).  This issue has relevance to the
so-called trans-Planckian problem (which in this BEC condensate
context becomes a trans-Bohrian problem). Fortunately it is known,
thanks to model calculations in field theories with explicit
high-momentum cutoffs, that the low energy physics of the emitted
radiation is largely insensitive to the nature and specific features
of the cutoff.  However, because of this high-frequency
``superluminal'' dispersion, one can have additional hopes of
detecting a signal because Hawking radiation could provide a new type
of instability, disrupting the classically stable configuration by
leading to a run-away production of phonons (see the discussion of
``black hole lasers'' in ~\cite{CT}).

To summarize what we have learned: The present note complements the
analysis by Garay \etal~\cite{garay}, in that it provides a rationale
for simple physical estimates of the Hawking radiation temperature
without having to solve the full Bogoliubov equations. Additionally,
the current analysis provides simple numerical estimates of the size
of the effect and identifies several specific physical mechanisms by
which the Hawking temperature can be manipulated: via the speed of
sound, the nozzle radius, the equation of state, and the degree of
flare-out at the throat.  In this manner, we have identified an
experimentally plausible configuration with a Hawking temperature of
order $70$ n K; to be contrasted with a condensation temperature of
the order of $90$ n K.

\emph{Acknowledgements:}
S.~Liberati wishes to thank B.L.~Hu and T.A.~Jacobson for illuminating
discussions and remarks. M.~Visser wishes to thank M.~Stone for
stimulating comments regarding the nozzle equation.


\end{document}